\documentclass[%
 aip,
 amsmath,amssymb,
 reprint,%
]{revtex4-1}

\usepackage{graphicx}
\usepackage{dcolumn}
\usepackage{bm}

\usepackage[utf8]{inputenc}
\usepackage[T1]{fontenc}
\usepackage{mathptmx}

\usepackage{color}

\begin{document}

\preprint{AIP/123-QED}

\title[Topological and geometric quantities]{Topological and geometrical quantities in active cellular structures}

\author{D. Wenzel}%
 \email{dennis.wenzel@tu-dresden.de.}
\affiliation{ 
Institute of Scientific Computing, TU Dresden, 01062 Dresden, Germany
}%
\author{S. Praetorius}%
\affiliation{ 
Institute of Scientific Computing, TU Dresden, 01062 Dresden, Germany
}%
\author{A. Voigt}
\affiliation{ 
Institute of Scientific Computing, TU Dresden, 01062 Dresden, Germany
}%
\affiliation{%
Dresden Center for Computational Materials Science (DCMS), TU Dresden, 01062 Dresden, Germany 
}%
\affiliation{%
Center for Systems Biology Dresden (CSBD), Pfotenhauerstr. 108, 01307 Dresden, Germany 
}%

\date{\today}

\begin{abstract}
Topological and geometrical properties and the associated topological defects find a rapidly growing interest in studying the interplay between mechanics and the collective behavior of cells on the tissue level. We here test if well studied equilibrium laws for polydisperse passive systems such as the Lewis's and the Aboav-Weaire's law are applicable also for active cellular structures. Large scale simulations, which are based on a multi phase field active polar gel model, indicate that these active cellular structures follow these laws. If the system is in a state of collective motion also quantitative agreement with typical values for passive systems is observed. If this state has not developed quantitative differences can be found. We further compare the model with discrete modeling approaches for cellular structures and show that essential properties, such as T1 transitions and rosettes are naturally fulfilled. 
\end{abstract}

\maketitle

\section{\label{sec:1}Introduction}

Although driven and active systems are far from equilibrium, they have been shown to share key features with passive systems. Examples are effective thermal behavior and time correlation functions in assemblies of cells, which behave as equilibrium glass transitions \cite{Berthieretal_NP_2013,Berthieretal_PRL_2009} or motility induced phase separation, which shares properties such as coarsening laws and statistical self-similarity with classical phase separation in binary systems \cite{Wittkowskietal_NC_2014,Specketal_PRL_2014}. We will here test if topological and geometric quantities, which are well studied for polydisperse assemblies in passive systems, as e.g. foams and froths \cite{Flyvbjerg_PRE_1993,Duplatetal_JFM_2011} or Ostwald ripening of minority phase domains after a rapid temperature quench \cite{Voorhees_JSP_1985,Moatsetal_PRB_2019}, are also valid for monolayers of cells. We will consider two empirical laws, the Lewis' law, originally proposed in studies of the epidermis of Cucumis \cite{Lewis_AR_1928}, it expresses the existence of correlation between area and number of neighbors (coordination number $q$) of a cell, and the Aboav-Weaire's law, with the original aim to understand the mechanism of growth of polycrystals \cite{Aboav_M_1970}, which states that cells with high (low) coordination number are surrounded by small (large) cells. In other words Lewis's law indicates how space is most likely to be filled by cells, whereas Aboav-Weaire's law gives the most probable correlation between neighboring cells. Both laws for space-filling cellular structures can be found in biological, geographical, mathematical and physical literature, see e.g. \cite{Chiu_MC_1995} for a review. Such topological properties and the associated topological defects find a rapidly growing interest in studying the interplay between mechanics and the collective behavior of cells on the tissue level \cite{Ladouxetal_NRMCB_2017}. Particularly for fully confluent epithelial tissues where cells are densely packed, various cell-based models have been developed to study epithelia tissue mechanics \cite{Farhadifaretal_CB_2007,Fletcheretal_BPJ_2014,Bietal_NP_2015,Yangetal_PNAS_2017,Bartonetal_PLOSCB_2017,Yanetal_arXiv_2018}. Besides these vertex models also phase field models that can represent a cells's shape and dynamics in great detail have been proposed \cite{Tjhung2012,Ziebert2012,Camleyetal_PNAS_2014,Loeber2015,Marth2015,Palmierietal_SR_2015,Marth2016,Camleyetal_JPD_2017}. These models, which fulfill essential properties of cellular structures naturally, will be the basis for the simulations in this paper. 

The paper is organized as follows: In Section \ref{sec:2} we briefly review the considered model, which is based on a multi phase field approach using one phase field variable per cell and a polar active gel theory within each cell. We further mention the numerical approach, which considers one cell per processor and thus shows parallel scaling properties independent of the number of cells. We will consider simulations with approximately 100 cells. In Section \ref{sec:3} the algorithm is used to analyze the collective behavior in various settings and topological and geometric properties are computed and compared with the equilibrium Lewis' and Aboav-Weaire's laws, and typical values obtained for passive systems. We thereby demonstrate the possibility to classify cellular systems according to their collective behavior. In Section \ref{sec:4} we draw conclusions.   

 \section{\label{sec:2}Model and Methods}

Each cell is modeled by a phase field active polar gel model \cite{Kruse2000,Kruse2004,Marth2015} and the cells interact via a short-range interaction potential \cite{Marth2016}. 
We consider $i = 1, \ldots, n$ phase field variables $\phi_i$ and polarization fields $\mathbf{P}_i$, for which the coupled evolution equations read 
\begin{equation*}\begin{split}\label{eq:evolution}
&\partial_t\phi_i + v_0\nabla\cdot\big(\phi_i\mathbf{P}_i\big) = \gamma \Delta \mu_i\,, \\
&\mu_i := \frac{\delta \mathcal{F}}{\delta\phi_i} = \frac{1}{Ca}\Big(\!-\epsilon\Delta\phi_i + \frac{1}{\epsilon}W'(\phi_i)\!\Big) \\
      &\qquad\qquad\quad + \frac{1}{Pa}\Big(\!\!-\frac{c}{2}\|\mathbf{P}_i\|^2 - \beta\nabla\cdot\mathbf{P}_i\!\Big) \\
      &\qquad\qquad\quad + \frac{1}{In}\Big(\!B'(\phi_i)\sum_{j\neq i}w(d_j) + w'(d_i) d_i^\prime(\phi_i) \sum_{j\neq i}B(\phi_j)\!\Big), \\
&\partial_t\mathbf{P}_i + \big(v_0\mathbf{P}_i\cdot\nabla\big)\mathbf{P}_i = -\frac{1}{\kappa}\mathbf{H}_i\,, \\
&\mathbf{H}_i := \frac{\delta \mathcal{F}}{\delta\mathbf{P}_i} = \frac{1}{Pa}\Big(\!\!-c\phi_i\mathbf{P}_i + c\|\mathbf{P}_i\|^2\mathbf{P}_i - \Delta\mathbf{P}_i + \beta\nabla\phi_i\!\Big),
\end{split}\end{equation*}
in $\Omega\times(0,T]$ for some simulation end time $T>0$ and a two-dimensional domain $\Omega$. We assume in the following periodic boundary conditions for $\Omega$. The model results as a $H^{-1}$ gradient flow (conserved evolution) of the energy $\mathcal{F}$ w.r.t. $\phi_i$ and an $L^2$ gradient flow (nonconserved evolution) w.r.t. $\mathbf{P}_i$ with 
\begin{equation*}\label{eq:energy}\begin{split}
  &\mathcal{F}[\{\mathbf{P}_i\},\{\phi_i\}] = \\ 
  &\sum_i \Big( \frac{1}{Ca}\int_\Omega \frac{\epsilon}{2}\|\nabla\phi_i\|^2 + \frac{1}{\epsilon}W(\phi_i)\,\text{d}\mathbf{x} \\
  &\quad + \frac{1}{Pa}\int_\Omega \frac{1}{2}\|\nabla\mathbf{P}_i\|^2 + \frac{c}{4}\|\mathbf{P}_i\|^2(-2\phi_i + \|\mathbf{P}_i\|^2) + \beta\mathbf{P}_i\cdot\nabla\phi_i\,\text{d}\mathbf{x} \\
  &\quad + \frac{1}{In}\int_\Omega B(\phi_i) \sum_{j\neq i} w(d_j)\,\text{d}\mathbf{x} \Big)
\end{split}\end{equation*}
considering the surface energy for the cell boundaries, a polar liquid crystal energy in the cells and interaction terms. The parameters $Ca$, $Pa$ and $In$ act as weightings between these contributions. The surface energy is a classical Ginzburg-Landau function with double-well potential $W(\phi) = \frac{1}{4}(\phi^2-1)^2$ and interface thickness $\epsilon$. Additional surface properties, such as bending, have shown to be of small impact \cite{Marth2016} and are therefore neglected here. The polar liquid crystal energy is of Frank-Oseen type and $c$ and $\beta$ are parameters controlling the deformation of the polarization fields $\mathbf{P}_i$ within the cell bulk and the anchoring on the cell interface, respectively. The value of $\beta$, modeling the polymerization rate of actin filaments will be of particular relevance in the following considerations due to its strong influence on the emergence of collective motion, see \cite{Loeber2015,Marth2016} for details. The interaction term considers $B(\phi_i)=\frac{3}{\epsilon \sqrt{2}} W(\phi_i) \approx \delta_{\Gamma_i}$ an approximation of the surface delta function for the cell boundary $\Gamma_i = \{ \mathbf{x}\in \Omega \;|\; \phi_i(\mathbf{x})= 0 \}$ and an approximation of an interaction potential $w(d_j)$ with signed distance function $d_j$ with respect to the zero-line (cell boundary $\Gamma_j$) of $\phi_j$. Activity is introduced in the evolution equations by a self-propulsion term, with velocity value $v_0$, which will also be of particular importance in the following. For more details we refer to \cite{Kruse2000,Kruse2004,Marth2015,Marth2016}. In previous studies a short range repulsive interaction was considered, with 
\begin{equation*} \label{eq:interaction}
w(d_j)=\exp\Big(-\frac{d_j^2}{\epsilon^2}\Big), \quad \text{with} \quad d_j(\phi_j) = -\frac{\epsilon}{\sqrt{2}}\ln\frac{1+\phi_j(\mathbf{x})}{1-\phi_j(\mathbf{x})}
\end{equation*}
the signed distance function computed from the equilibrium $\tanh$-profile of the phase field function $\phi_j$. However, other forms are possible as well, such as short range adhesive interactions or even more realistic effective interaction potentials, as e.g. experimentally determined in \cite{Hartmannetal_NP_2018} and shown to be essential for local cellular order. Here our approach differs from other phase field studies, e.g. \cite{Camleyetal_PNAS_2014,Loeber2015}, where simply the overlap of two cells is penalized by an energy term proportional to $\sum_{j\neq i} \int_\Omega \phi_i^2 \phi_j^2 \, d \mathbf{x}$ and/or adhesion to the cell boundaries is promoted by a term proportional to $- \sum_{j\neq i} \int_\Omega |\nabla \phi_i|^2 |\nabla \phi_j|^2 \, d \mathbf{x}$.

In order to solve the equations numerically we consider a finite element implementation which scales with the number of cells $n$. This is achieved by considering each cell on a different processor and various improvements to reduce the communication overhead to deal with the cell-cell interactions, see \cite{Vey2007,Witkowski2015,Ling2016,Marth2016b,Praetoriusetal_NIC_2017} for detail.   

Previous studies of the model were concerned with dilute monodisperse systems and the emergence of collective motion \cite{Marth2016,Praetoriusetal_NIC_2017}. We here consider densely packed disperse systems where various model parameters are varied to analyze the effect on topological and geometric quantities of the active cellular structures.

 \section{\label{sec:3}Results}

\begin{figure}[h]
	\centering
	\includegraphics[width = 0.45\textwidth]{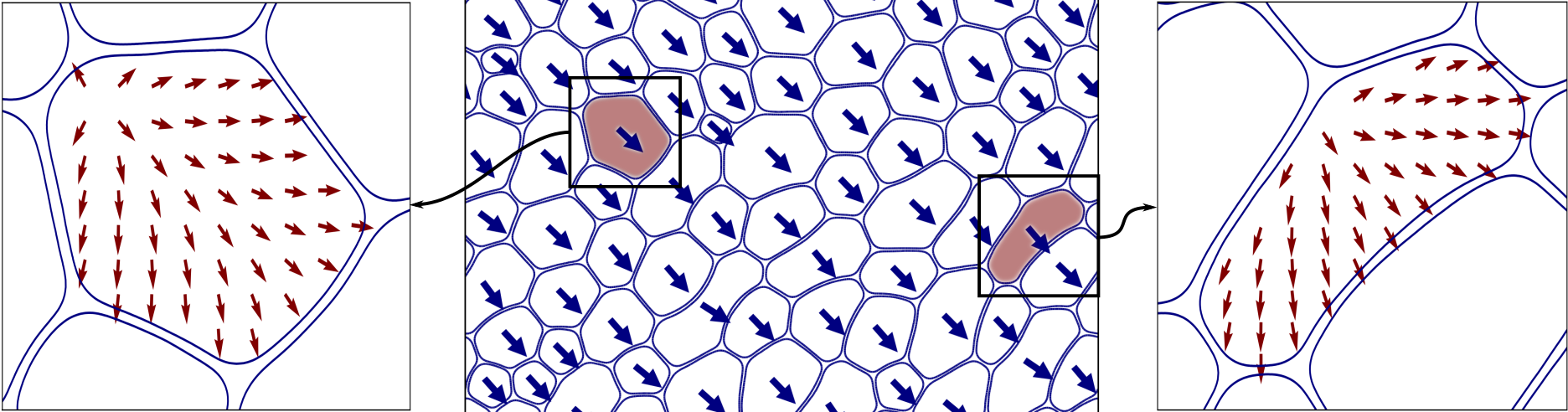}
	\caption{Examples of $\hat{\mathbf{P}}_i$ (middle) as average of $\mathbf{P}_i$ (left,right).}
	\label{fig:0}
\end{figure}

In the following we consider for each cell the average polarization defined by
\begin{equation*}
	\label{eq:p_average}
	\hat{\mathbf{P}}_i := \dfrac{\int_\Omega \mathbf{P}_i \left(\phi_i +1 \right)\text{d}\mathbf{x} }{\int_\Omega \left(\phi_i +1 \right)\text{d}\mathbf{x} }
\end{equation*}
and visualized in Figure \ref{fig:0}. Shown are the zero-level lines of $\phi_i$ (cell boundaries of each cell). The red arrows in the pictures left and right visualize $\mathbf{P}_i$ for two exemplary cells while the blue arrows in the middle frame depict $ \hat{\mathbf{P}}_i$ for all shown cells.

\begin{figure}[h]
  \centering
  \includegraphics[width = 0.45\textwidth]{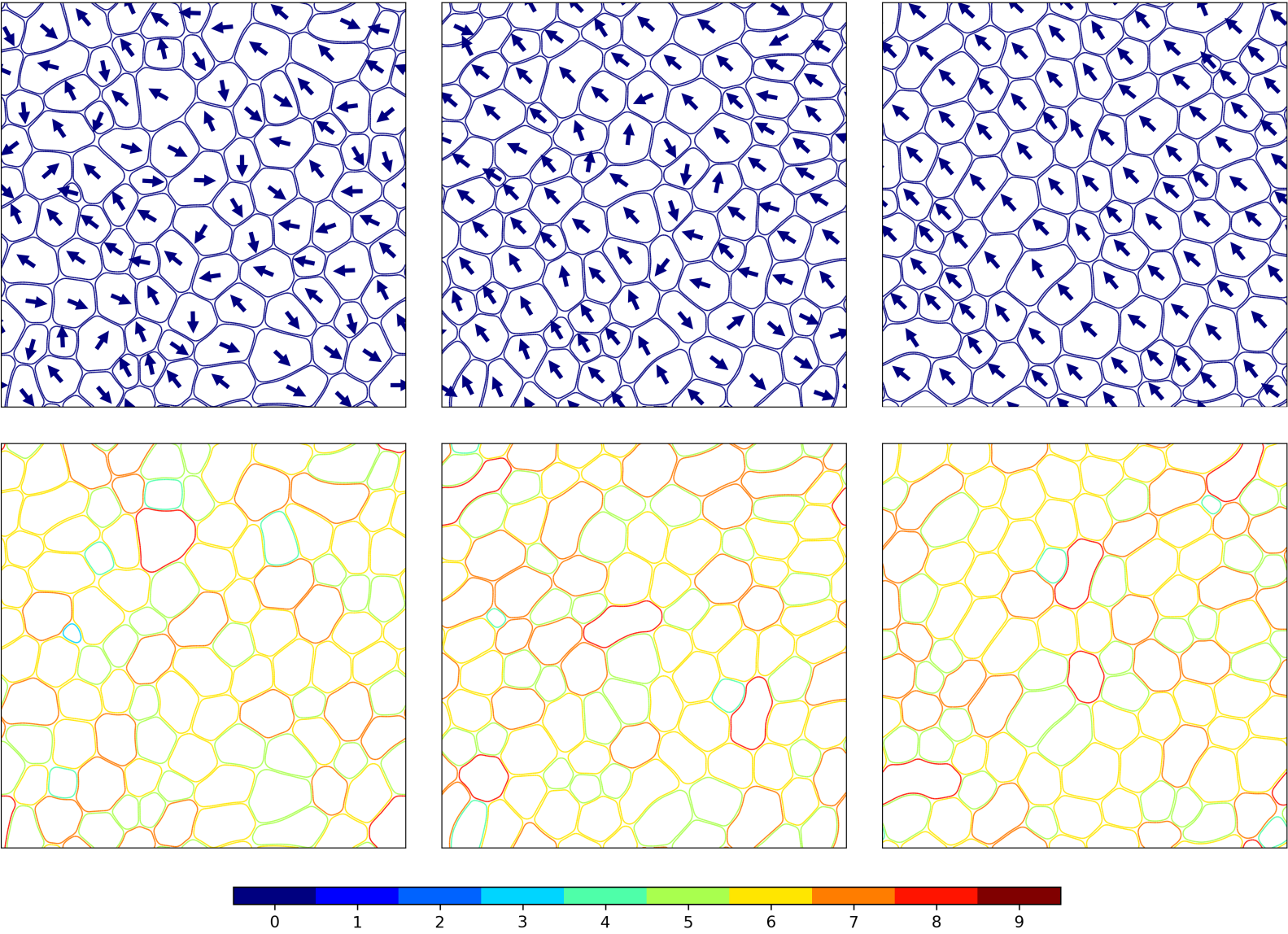} 
  \caption{Snapshots of the evolution for $t = 60, 225, 800$ from left to right. The simulation is done with $\beta = 0.3$, $v_0=2.5$ and a random initialization of the polarization field. The top row visualizes the emergence of collective motion. The bottom row shows the number of neighbors at the corresponding time instances with the color code given below.}
  \label{fig:1}
\end{figure}

Figure \ref{fig:1} shows snapshots of the evolution of the cellular structure with arrows indicating $ \hat{\mathbf{P}}_i$ for all cells. We in addition color code the number of neighbors (coordination number $q$), which is shown for the same snapshots. 

\begin{figure}[h]
  \centering
  \includegraphics[width = 0.45\textwidth]{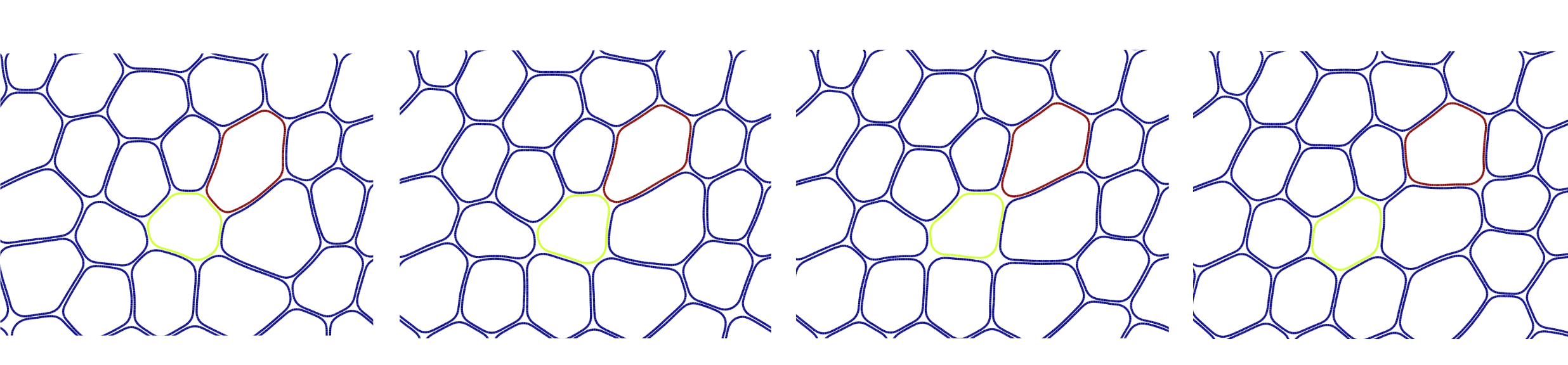} $\;$
 \caption{Typical T1 transition. The color highlights the involved cells in the topological change which are in contact with each other at the beginning of the T1 transition.}
  \label{fig:2a}
\end{figure}

\begin{figure}[h]
  \centering
  \includegraphics[width = 0.45\textwidth]{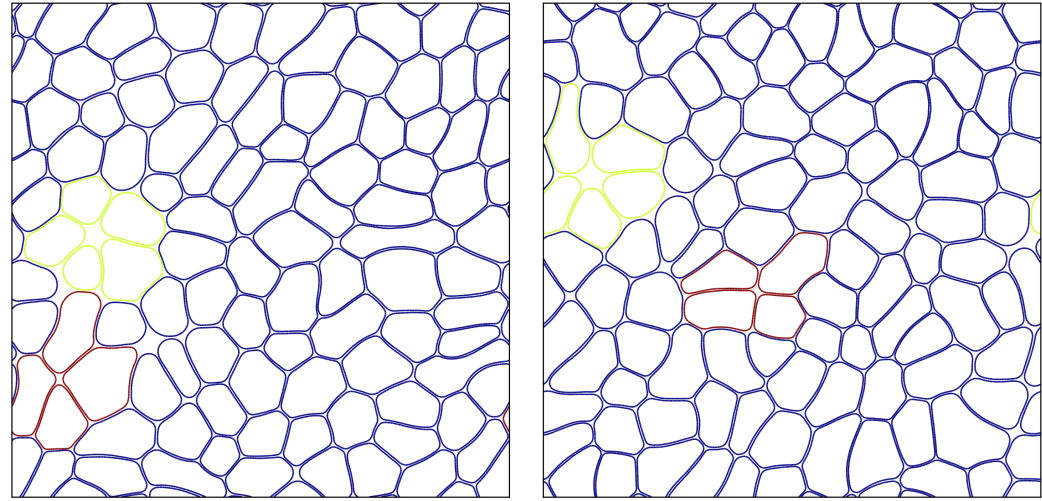} 
 \caption{Examples for rosettes found in the simulations. The color highlights "vertices" with four (red) or five (green) cells.}
  \label{fig:2b}
\end{figure}

We consider an initial size distribution with a mean cell area determined by the number of cells and a variance of $\sigma = 3$. In particular the initial condition of the phase field variables is given by a dense packing of rectangular cells aligned in a brick wall pattern. Due to the phase field approach with $\epsilon > 0$ a space-filling structure is not possible, we instead consider an area fraction of $0.95$ achieved by a shrinkage of each cell, which explains the visible empty space in Figures \ref{fig:1} - \ref{fig:2b}. We vary the parameter $\beta$, which models the anchoring of the polarization field at the cell boundaries, and the parameter $v_0$, which models the self-propulsion. When studying the influence of $\beta$ we choose $v_0=2.5$ and in the study of $v_0$ we fix $\beta =0.3$. All other parameters are fixed over all simulations, see Table \ref{tab:1} for the considered values.
\begin{table}
\begin{tabular}{|c|c|c|c|c|c|c|c|c}
\hline
$\Omega$ & $\epsilon$ & $\gamma$ & $\kappa$ & $Pa$ & $Ca$ & $In$  & $c$ \\ \hline \hline
$[0,100] \times [0,100]$ & 0.15           & 1                & 1             & 1        & 0.025 & 0.05 & 1 \\ \hline
\end{tabular}
\caption{Numerical parameters used in all simulations.} \label{tab:1}
\end{table}

In contrast to vertex models, e.g. \cite{Farhadifaretal_CB_2007}, where topological transitions have to be incorporated by hand, the multi phase field approach deals with these transitions automatically, see Figure \ref{fig:2a} for a typical T1 transition in the evolution process. Also points where four or more cells are in contact with each other, so called rosettes, have to be explicitly enforced within vertex models \cite{Yanetal_arXiv_2018}, and occur naturally within the multi phase field approach, see Figure \ref{fig:2b} for examples. Such rosettes have been identified as crucial in development, disease and physiology, see e.g. \cite{Hardingetal_Dev_2014,Trichasetal_PLOSB_2012,Wippoldetal_AJNR_2006}.

\begin{figure}[h]
  \centering
  \includegraphics[width = 0.45\textwidth]{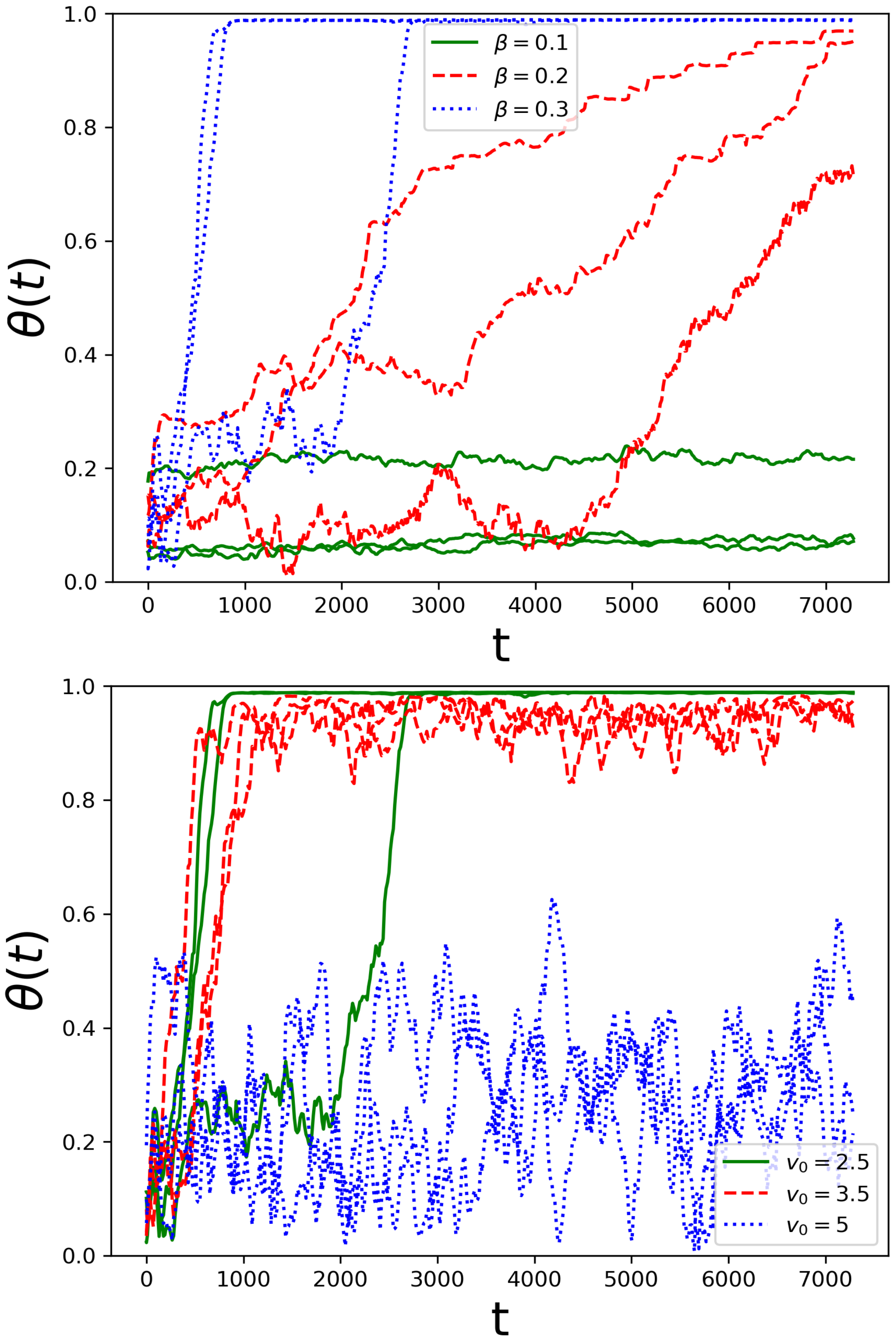} 
 \caption{Translational order parameter $\theta(t)$ indicating collective motion for different values of $\beta$ with fixed $v_0=2.5$ (top) and different values of $v_0$  with fixed $\beta=0.3$ (bottom). The time is considered in non-dimensional units.}
  \label{fig:3}
\end{figure}

We are now concerned with the emerging behavior. Depending on the parameters $\beta$ and $v_0$ the cellular structure converges to a state where all cells move in the same direction or remains within the considered time $T$ in a chaotic state. We quantify this evolution according to \cite{Loeber2015,Alaimo_2016} by computing the translational order parameter
\begin{equation*}
\theta(t) = \frac{1}{n} \Big\| \sum_i \hat{\mathbf{v}}_i(t) \Big\|
\end{equation*}
with $\hat{\mathbf{v}}_i(t)$ the unit velocity vector of cell $i$ at time $t$. It is $1$ if all cells move in the same direction and $0$ if all cell velocities are independent. In particular we use the threshold $0.9<\theta$ to classify a state of global collective motion. Note that this alignment of all particles in an active system is sometimes also referred to as a state of flocking, see \cite{Toner1998,TONER2005170}. The cell velocity is computed from the center of mass of each cell at adjacent time instances. Its direction corresponds to the direction of $\hat{\mathbf{P}}_i$. Figure \ref{fig:3} (top) shows the evolution of the order parameter $\theta$ for different initial conditions and different $\beta$ with fixed $v_0=2.5$. For $\beta = 0.3$ a state of collective motion is reached immediately, whereas for $\beta = 0.1$ this state is never reached within the considered time $T$ and one could conclude that it will probably never be reached. For $\beta = 0.2$ it takes a long time before collective motion can be observed. Depending on the initial conditions it might not even be reached within the considered time \cite{comment}. Figure \ref{fig:3} (bottom) on the other hand shows the evolution of $\theta$ with different values of $v_0$ but fixed $\beta=0.3$. In comparison to $v_0=2.5$ we observe that for the slightly increased value $v_0=3.5$ the system still reaches a collective state after relatively short time although oscillations in the translational ordering are strengthened. Upon increasing the activity further to $v_0=5.0$ no simulation showed collective motion in the considered time $T$ and oscillations in $\theta$ are increased even more. These findings are in qualitative agreement with results obtained for more dilute systems \cite{Praetoriusetal_NIC_2017}. 

\begin{figure}[h]
  \centering
  \includegraphics[width = 0.45\textwidth]{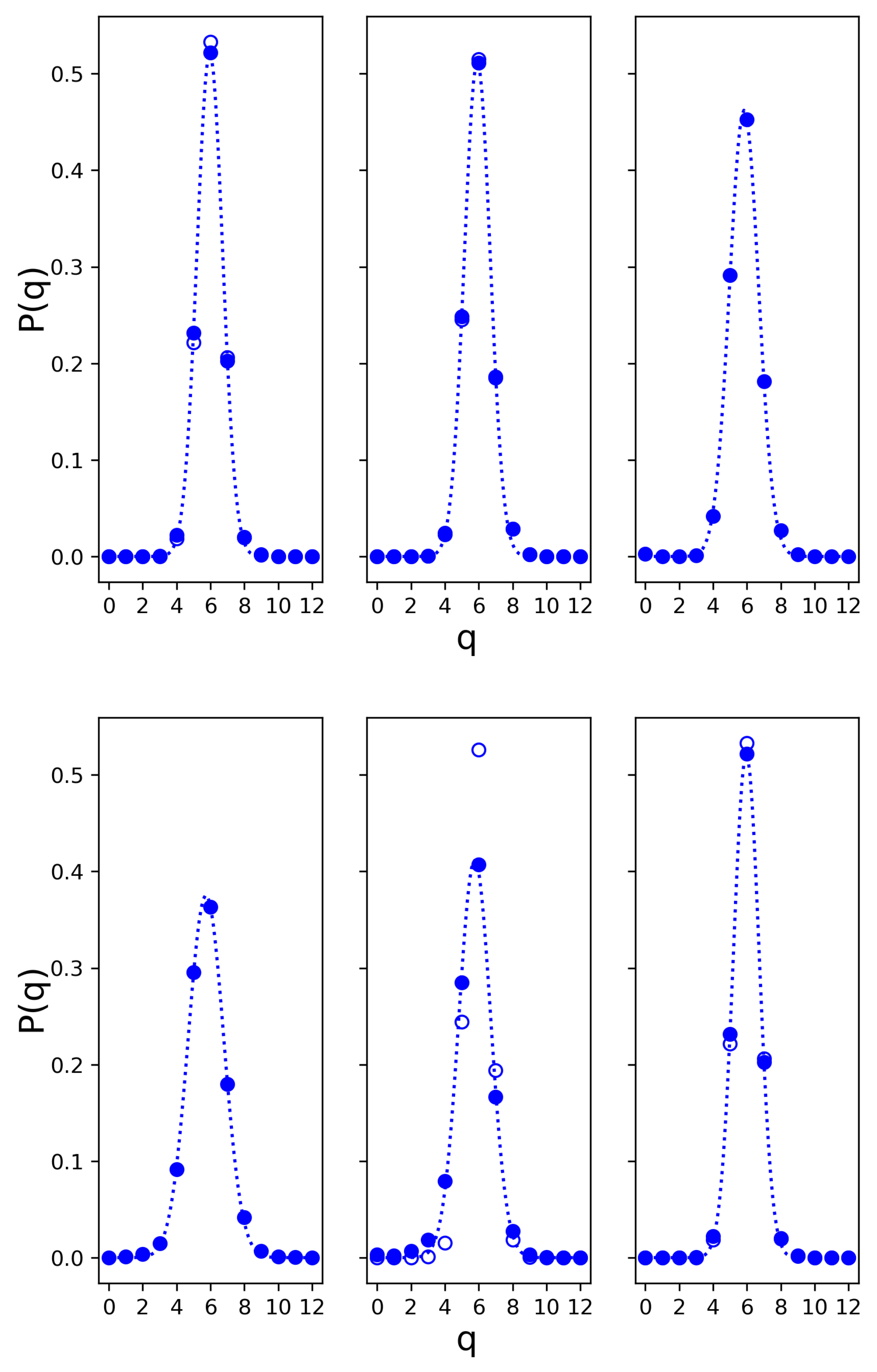} 
 \caption{Coordination number probability for $\beta = 0.1, 0.2, 0.3$ with fixed $v_0=2.5$ (top) and $v_0=2.5,3.5,5.0$ with $\beta=0.3$ (bottom), from left to right. Shown is the average of the whole time evolution and all considered samples (closed symbols and fit) and average over the time, where collective motion is already reached ($0.9 < \theta$) (open symbols).}
  \label{fig:4}
\end{figure}

In order to characterize the topological domain structures during the evolution we consider the coordination number probability of the cells, $P(q) = N(q) / N$, where $N(q)$ is the number of cells with $q$ nearest neighbor cells and $N = \sum_q N(q)$. Figure \ref{fig:4} shows $P(q)$ averaged over the whole time interval and the three different simulations for each $\beta$ (top) and each $v_0$ (bottom). $P(q)$ is symmetric and centered at $q = 6$. This is true for all considered $\beta$ and $v_0$. Studying the dependence on $\beta$ we observe that $P(q)$ is time-independent for $\beta=0.1$ and $\beta=0.3$. For $\beta = 0.2$ it evolves towards the functional form for $\beta = 0.3$ if the state of collective motion is reached, see additional open symbols in Figure \ref{fig:4}, which are averaged only over the time, where collective motion is reached.. For varying values of $v_0$ we observe a time-independent behavior for all simulations. $P(q) \neq 6$ corresponds to cellular structures with topological defects. Depending on $\beta$ and with fixed $v_0=2.5$ we obtain $\mu_2 = \sum_q (q - 6)^2 P(q) \approx \{1.05; 0.94; 0.76\}$ for $\beta = 0.1; 0.2; 0.3$, respectively, for the variance of $P(q)$. Considering for $\beta = 0.2$ only the time, where collective motion is reached, which is identified by $0.9 < \theta$, we obtain $\mu _2 = 0.75$. Depending on $v_0$ and with fixed $\beta=0.3$ we obtain $\mu_2 \approx \{0.76, 0.77, 0.86\}$ for $v_0 = 2.5; 3.5; 5.0$, respectively. $P(q)$ and $\mu_2$ have been computed for various passive systems, experimental and theoretical. They are all symmetric and centered at $q = 6$. And also the variance values, if collective motion is reached, are close to these measured data for passive systems, see e.g. \cite{Sireetal_JPI_1995}, where $\mu_2 = 0.64$ has been reported. The coordination number probability of an active cellular structure thus seems to behave as in passive systems if it is in the state of collective motion, and quantitatively differs if not. 

\begin{figure*}[th]
  \centering
  \includegraphics[width = \textwidth]{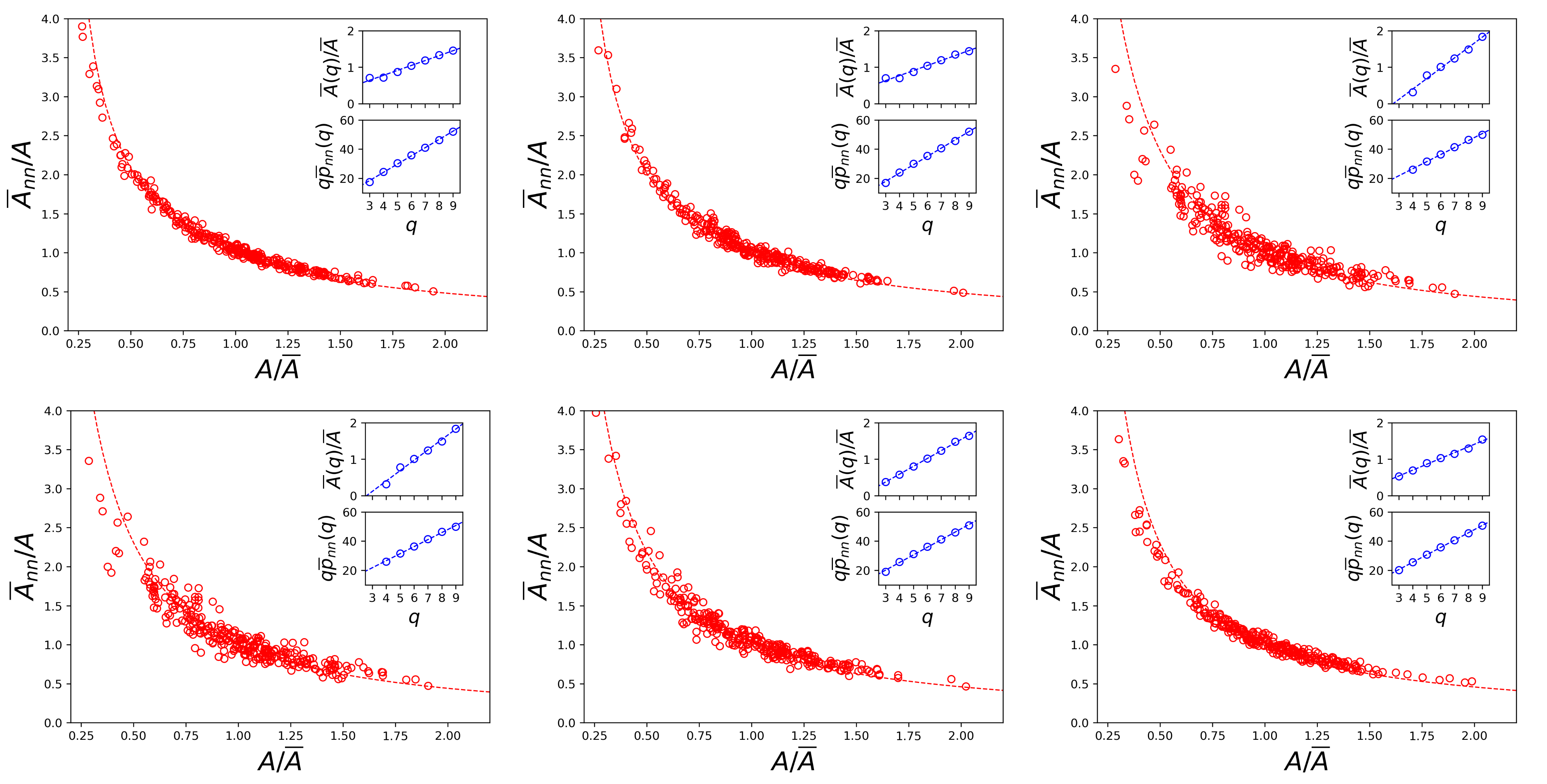} 
 \caption{Normalized average area of nearest neighbor cells, $\bar{A}_{nn}/A$ vs. $A/\bar{A}$ for $\beta = 0.1, 0.2, 0.3$ with fixed $v_0=2.5$ (top row) and $v_0=2.5,3.5,5.0$ with fixed $\beta=0.3$ (bottom row) from left to right, together with a fit according to eq. \eqref{eq:1}. Inset (top) shows $\bar{A}(q)/A$ vs. coordination number $q$, corresponding to Lewis's law. The line shows a linear fit through the data. Inset (bottom) shows the average coordination number of nearest neighbor cells of $q$-coordinated cells vs. coordination number $q$, corresponding to Aboav-Weaire's law. The line shows a linear fit through the data.}
  \label{fig:5}
\end{figure*}

Lewis's law \cite{Lewis_AR_1928} states that cells with a coordination number $q = 6$ tend to have a size equal to the average cell size, and that cells that are larger (smaller) than the average cell size tend to have a coordination number larger (smaller) than six and reads
\begin{equation}
\frac{\overline{A}(q)}{\overline{A}} = \alpha (q - 6) + \gamma
\end{equation}
with $\overline{A}(q)$ the average area of $q$-coordinated cells, $\overline{A}$ the average cell area and $\alpha$ and $\gamma$ scalar fitting parameters. The results are shown in Inset (top) of Figure \ref{fig:5}. Depending on $\beta$ and with fixed $v_0=2.5$ we obtain $(\alpha, \gamma) = (0.14, 1.05); (0.14, 1.05); (0.28, 0.97)$ for $\beta = 0.1;0.2; 0.3$, respectively. If for $\beta = 0.2$ only the times, where collective motion is already reached ($0.9 < \theta$), is considered, one obtains $(\alpha, \gamma) = (0.19, 0.99)$. Depending on $v_0$ and with fixed $\beta=0.3$ on the other hand we obtain $(\alpha, \gamma) = (0.28, 0.97); (0.21, 1.02); (0.16, 1.02)$ for $v_0 = 2.5; 3.5; 5.0$, respectively. We observe that the values for situations with collective motion are again in excellent agreement with data in \cite{Sireetal_JPI_1995}, where $0.20 \leq \alpha \leq 0.25$ and $0.95 \leq \gamma \leq 1.05$ has been reported.

Aboav-Weaire's law \cite{Aboav_M_1970} describes topological correlation between the coordination number of a cell, $q$, with the average coordination number of its nearest neighbor cells $\overline{p}_{nn}(q)$ and reads
\begin{equation}
q \overline{p}_{nn}(q) = (6 - \zeta)(q - 6) + \eta
\end{equation}
with scalar fitting parameters $\zeta$ and $\eta$. The results in Inset (bottom) of Figure \ref{fig:5} show the suggested linear behavior. Depending on $\beta$ and with fixed $v_0=2.5$ we obtain $(\zeta, \eta) = (0.36, 35.35);(0.32,34.97);(1.18,36.22)$ for $\beta = 0.1; 0.2; 0.3$, respectively, and for the times of collective motion ($0.9 < \theta$) for $\beta = 0.2$, $(\zeta, \eta) = (1.20, 35.93)$. Depending on $v_0$ and with fixed $\beta=0.3$ we obtain $(\zeta, \eta) = (1.18, 36.2);(0.79,35.76);(0.96,35.62)$ for $v_0 = 2.5; 3.5; 5.0$, respectively. Again the values for collectively moving cellular structures are in quantitative agreement with those measured in passive systems, see \cite{Sireetal_JPI_1995}, where $(\zeta, \eta) = (1.10, 36.64)$ have been reported. The linearity of the Aboav-Weaire's law has also been found theoretically \cite{Godrecheetal_PRL_1992} and experimentally \cite{Mombachetal_JPD_1990}. 

Both laws, the Lewis's law and the Aboav-Weaire's law, also lead to a correlation between the area of a cell, $A$, and the average area of its nearest neighbors, $\overline{A}_{nn}$, see Figure \ref{fig:5}. Cells which are larger (smaller) than the average cell size are mostly surrounded by nearest neighbor cells that are smaller (larger) in size. All three cases qualitatively show this anticorrelation and also fulfill the proposed functional form for this relation, 
\begin{equation}
\label{eq:1}
f(x) = \frac{1}{x} \Big( 1 + \frac{\alpha ^2 \mu_2 - \zeta \alpha (x -1)}{6 \alpha + (x -1)} \Big)
\end{equation}
with $x = A / \bar{A}$ and $f(x) = \bar{A}_{nn} / A$ and the fitting data $\alpha$ from Lewis's law, $\zeta$ from Aboav-Weaire's laws and $\mu_2$ the variance of $P(q)$. The functions are plotted in Figure \ref{fig:5}. The law results from maximum entropy theory for random two-dimensional cellular structures \cite{Peshkinetal_PRL_1991,Seuletal_PRL_1994}. 

In summary, we obtain qualitative agreement with equilibrium topological and geometrical relations widely found in passive systems, independent of the used parameters and the macroscopic state of the active cellular structure and even quantitative agreement with typical values for Lewis's law, Aboav-Weaire's law and the combined functional form in eq. \eqref{eq:1} for passive systems, within the state of collective motion. To further confirm these results we classify the whole data set (all $\beta$, all $v_0$, all initial conditions) according to the value of the translational order parameter $\theta$, with $\theta < 0.3$ (chaotic regime), $0.3 \leq \theta \leq 0.9$ (intermediate regime) and $0.9 < \theta$ (collective motion), we obtain the values in Table \ref{tab:2}.
\begin{table}
\begin{tabular}{cccccc}
\hline
 & & \multicolumn{2}{c}{Lewis's law} & \multicolumn{2}{c}{Aboav-Weaire's law} \\ \hline \hline 
 & $\mu_2$ &$\alpha$ & $\gamma$ & $\zeta$ & $\eta$ \\ 
 $\theta < 0.3$ & 0.98 & 0.14 & 1.04 & 0.64 & 35.25 \\ 
 $0.3 \leq \theta \leq 0.9$ & 0.88 & 0.17 & 1.03 & 0.69 & 35.40 \\ 
 $0.9 < \theta$ & 0.78 & 0.26 & 0.97 & 1.11 & 36.28 \\ 
 Sire et al. \cite{Sireetal_JPI_1995} & 0.64 & [0.20, 0.25] & [0.95,1.05] & 1.10 & 36.64 \\  \hline
\end{tabular}
\caption{Variance of coordination number probability $P(q)$ and linear fitting parameters for topological and geometric laws over all simulations, classified according to translational order parameter $\theta$.} \label{tab:2}
\end{table}

Our investigations indicate that also in active cellular structures, if they are collectively moving, the cells are arranged in space to maximize the configurational entropy. In this study we only varied the parameters $\beta$, which controls the anchoring of the polarization fields $\mathbf{P}_i$ on the cell interface, see \cite{Marth2015,Marth2016}, and $v_0$, which induces activity of the system. All other parameters are kept constant and further studies, e.g. on the influence of the stiffness of the cell boundaries or the interaction potential, are postponed to future work. However, the classification according to the translational order parameter $\theta$ suggests to obtain similar results also in these situations.

 \section{\label{sec:4}Conclusions}

The used multi phase field active polar gel model can be considered as a minimal model of active cellular structures. It is based on active driving, cell deformation and force transmission through interactions at the cell-cell interfaces. The collective dynamics in the cellular structures are not simply the result of many individually moving cells, but results from coordinated movement of interacting cells. It could not be expected that laws, which are derived for passive systems, are applicable for such active structures. However, our results clearly indicate that topological and geometric relations also hold in active systems and established laws, such as the Lewis's law and the Aboav-Weaire's law are fulfilled if the active system is in a collectively moving state. Using the translational order parameter with $0.9 < \theta$ as an indicator for collective movement all simulation results fulfill these laws and agree quantitatively with typical results for passive systems. Below this threshold the results deviate from these values.  

The proposed multi phase field model shows various advantages if compared with active vertex models \cite{Bartonetal_PLOSCB_2017} and the considered numerical approach allows to simulate a reasonable number of cells to analyze characteristic tissue topologies. However, for its applicability in tissue mechanics, various model extensions, e.g. towards cell growth and cell division are required. On the other side, even simplified approaches of the multi phase field model have recently been used to demonstrate the emergence of macroscopic nematic liquid crystal features of tissue \cite{Muelleretal_arXiv_2018}. The corresponding topological nematic defects have shown to be essential for cell death and extrusion \cite{Sawetal_N_2017}. To analyses the relation between the dynamics in cellular structures and such macroscopic features and the resulting tissues mechanics, still has to be explored. The multi phase field approach is a valuable framework that can incorporate a broad range of submodels, in our case an active polar gel model. Other possibilities are couplings with biochemistry \cite{Marth2014,Camleyetal_PRE_2017} or hydrodynamics \cite{Marth2016}.    

\begin{acknowledgments}
We acknowledge computing resources provided at JSC under grant HDR06.
\end{acknowledgments}

\nocite{*}
\bibliography{short_names,references}

\begin{thebibliography}{50}
\bibitem [{com()}]{comment}%
  \BibitemOpen
  \href@noop {} {\bibinfo  {journal} {The final time and the number of
  performed simulations are a compromise between predictability and economic
  use of computing resources}\ }\BibitemShut {NoStop}%
\end{thebibliography}

\end{document}